\documentclass[12pt,a4paper]{article}
\pdfoutput=1

\usepackage{graphicx}
\usepackage{bm}
\usepackage{amsmath}
\usepackage{amssymb,xfrac}
\usepackage[bbgreekl]{mathbbol}
\usepackage{cite,color}

\begin{document}

\pagestyle{plain}

\begin{center}
~

\vspace{1cm} {\large \textbf{
A Lattice Inspired Model for Monopole Dynamics
}}

\vspace{1cm}

Amir H. Fatollahi

\vspace{.5cm}

{\it Department of Physics, Faculty of Physics and Chemistry, \\ 
Alzahra University, Tehran 1993891167, Iran}

\vspace{.3cm}

\texttt{fath@alzahra.ac.ir}

\vskip .8 cm
\end{center}

\begin{abstract}
The site-reduction of U(1) lattice gauge theory 
along the spatial directions is used to model the monopole dynamics. 
The reduced theory is that of the angle-valued coordinates on 
the discrete worldline. 
Below the critical coupling $g_{c}=1.125$ and temperature $T_c=0.335$ the model 
exhibits a first order phase transition. 
It is argued that the phase structure matches with the proposed role for 
magnetic monopoles in the confinement mechanism based on the dual Meissner effect.
\end{abstract}

\vspace{1cm}



\newpage


The theoretical \cite{lattice,kogut,banks,savit,guth,spencer,jaffe,polya1,polya2,thooft1} and lattice 
simulation \cite{creutz,nauen,bha,moria,degran,wiese,arnold,langfeld} studies strongly suggest that 
the 4D Abelain U(1) gauge theory exhibits a phase transition between 
confined and Coulomb phases. According to the scenario based on the dual Meissner 
effect in superconductors, the magnetic monopoles play a distinguished role in the phase transition 
\cite{banks,savit,nambu,mandelstam,thooft2}; for details see \cite{ripka}. 
At the strong coupling limit where the monopoles have a tiny mass and high density, the 
motion of monopoles around the electric field fluxes prevents them to spread over 
the space, leading to the confinement of electric charges. 
Instead, at the small coupling limit where the monopoles are massive and dilute, the field fluxes originated 
from electric charges are likely to spread, leading to the Coulomb's law. 
Supposedly there are critical coupling and temperature $g_c$ and $T_c$ at which the transition between 
two phases occurs. The lattice simulations suggest $g_c\simeq 1$ 
\cite{creutz,nauen,bha,moria,degran,wiese,arnold,langfeld}.

By now the best known framework to study the above mentioned phase transition is the lattice formulation 
of the U(1) theory, in which the gauge fields are treated as compact angle variables. The pure U(1) 
action defined on
the Euclidean lattice is given by \cite{lattice,kogut}
\begin{align}\label{01}
S_\mathrm{lattice}=\frac{1}{2g^2}\sum_{\mu,\nu}\sum_{\vec{n}}
\left( \cos(f_{\vec{n},\mu\nu})-1\right)
\end{align}
in which the basic object assigned to each lattice plaquette of size ``$\,a\,$" is defined by
\begin{align}\label{02}
f_{\vec{n},\mu\nu}:=
 a(A_{\vec{n},\mu}+ A_{\vec{n}+\hat{\mu},\nu}-A_{\vec{n}+\hat{\nu},\mu}-A_{\vec{n},\nu})
\end{align}
with $A_{\vec{n},\mu}$ as the gauge field at lattice site $\vec{n}$ in direction $\mu$, and $\hat{\mu}$ 
as the unit-vector. It is assumed $-\pi\leq a\,A \leq\pi$ \cite{lattice}.
In the weak coupling limit $g\ll 1$ the dynamics is essentially captured by the 
configurations $a A\ll 1$, by which (\ref{01}) is reduced to the continuum theory
$-\frac{1}{4g^2} \int d^4 x\, F_{\mu\nu}^2$
with $F_{\mu\nu}=\partial_\mu A_\nu-\partial_\nu A_\mu$.
While in the ordinary formulation of the U(1) theory on the continuum the monopoles are absent,
thanks to the compact nature of the gauge fields, 
the lattice formulation contains the monopoles 
\cite{banks,polya2,degran,fromar}. The studies based on the lattice formulation have provided 
clear evidences for the role of monopoles in the above mentioned phase transition 
\cite{degran,wiese,kerler,zach,barber}.

In the present note a model for the monopole dynamics is proposed in which the position variables are 
angle-valued in the very sense of gauge fields in lattice formulation. In particular, we consider
the following as the action which captures the essential dynamics by the monopoles
\begin{align}\label{03}
S_\mathrm{mon.}=\frac{1}{g^2}
 \sum_{i=1}^d\sum_n \left(\cos\frac{x^i_{n+1}-x^i_{n}}{R}-1\right)
\end{align}
in which `$n$' represents the dependence on the discrete imaginary-time with spacing `$a$', and 
`$R$' determines the extent of the compact position variable, $-\pi R\leq x \leq\pi R$.
The above model is in fact the sum of $d$ independent copies of the 1D XY model of magnetic systems. 
The close relation between lattice gauge theories and
spin systems was recognized from the first appearance of these theories \cite{lattice,kogut}, 
and has been used widely for better understanding the gauge theory side. 
The model was originally introduced as a spin-chain interpretation of the world-line in \cite{spchfath}, 
where the possible application of the model to monopole dynamics was discussed; 
see also \cite{pvfath,lost} for other related works.
The model may be regarded as a continuation of the agenda by which, it is insightful 
to demand that the coordinates and fields have similar characters 
\cite{dualfath,matone, dualfermion}. In fact, as far as the compact variables are concerned, 
the relation between the lattice gauge model and the above particle dynamics may be regarded 
as the interchange of the roles between dimensionless compact variables 
\begin{align}\label{04}
a\, A^i\longleftrightarrow   x^i/R
\end{align}
As another ground for the above relation, it is worthwhile to mention that 
a very similar interchange between gauge fields and coordinates is 
in fact originated by the T-duality of string theory. 
In particular, it is understood that upon certain conditions, the compactified version of 
a theory on radius $R_1$ is equivalent to a dual one compactified on radius $R_2$,
provided that two radii satisfy $R_1\,R_2 \simeq\alpha'$ \cite{tasi}.
Accordingly, the dynamics of $p$-dimensional 
non-perturbative solitonic objects emerged in the dual model, the so-called D$p$-branes, 
is captured by the dimensional reduction of gauge theories, interpreting the gauge fields 
as D$p$-brane transverse coordinates \cite{tasi,9510017,9510135}.
The gauge fields of original model and 
the position of emerged D-brane in the dual model satisfy \cite{tasi}
\begin{align}\label{05}
	R_1\, A = x/R_2
\end{align}
In the very same sense, following compactification on radii $R_1=a$  and $R_2=R$, 
the interchange (\ref{04}) takes place after dimensional reduction of the lattice action 
(\ref{01}).

In the first place let us consider the weak coupling limit $g\ll 1$ of the action (\ref{03}),
by which the dynamics is essentially captured by the configurations $x/R\ll 1$, leading to 
\begin{align}\label{06}
S_\mathrm{mon.} \simeq -\frac{m}{2a}\sum_{i,n} (x_{n+1}^i-x_n^i)^2 
\end{align}
representing the Euclidean action in the time-sliced form \cite{wipf} of a particle with mass 
\begin{align}\label{07}
m=\frac{a}{g^2R^2}
\end{align}
The above is consistent with the expectation that the mass of magnetic monopoles 
is proportional to $1/g^2$. First let us consider the thermodynamics at weak coupling 
by (\ref{06}).
As the model is the direct sum of $d$ copies of one-dimensional case, 
it is sufficient to consider only the $d=1$ model.
The one-particle partition function at temperature $T=\beta^{-1}$ 
for a particle of mass $m$ in the box $[-L/2,L/2]$ is represented in 
terms of the Euclidean action $S_E$ as \cite{wipf}
\begin{align}\label{08}
Z_1= \int_{-L/2}^{L/2}\, \prod_{n=0}^{N-1} 
\sqrt{\frac{m}{2\pi a}}\,dx_n~ e^{S_E}
\end{align}
with $a=\beta / N$ as the time-slice parameter. 
The representation (\ref{08}) is supplemented by the periodic 
condition $x(0)=x(\beta)$ (\textit{i.e.} $x_0=x_N$). 
By the action (\ref{06}) at weak coupling limit $g\ll 1$, setting 
$L=2\pi R$ and $m$ by (\ref{07}) after the change $y=x/(gR)$, one finds
\begin{align}\label{09}
Z_1= \int_{-\pi/g}^{\pi/g}\, \prod_{n=0}^{N-1} 
\frac{dy_n}{\sqrt{2\pi}}\, \exp\left(-\sum_{n=0}^{N-1}(y_{n+1}-y_n)^2\right)
\end{align}
We see that the volume of the box is practically $V_g=2\pi/g$. 
Due to the limit $g\ll 1$ the integrals can be considered as the Gaussian ones, 
leading to the well known result for free particles \cite{wipf}
\begin{align}\label{10}
Z_1= \sqrt{\frac{2\pi a}{\beta}}V_g
\end{align}
For the monopoles it is expected that the mass does vary with the 
gauge coupling. So, by means of the free energy 
$A_1=-T\ln Z_1$, we define the mass-conjugate variable 
\begin{align}\label{11}
M:= - \frac{\partial A_1}{\partial m} 
 = \frac{T}{Z_1} \,\frac{\partial Z_1}{\partial m}
\end{align}
for which we find 
\begin{align}\label{12}
M=\frac{T}{2\, m} \propto \langle \dot{x}^2 \rangle
\end{align}
By above the isothermal $M$-$m$ curves are simply the reciprocal functions (see Fig.~1), and evidently
they do not exhibit any kind of phase transition. 
\begin{figure}[!ht]
	\begin{center}
		\includegraphics[scale=0.7]{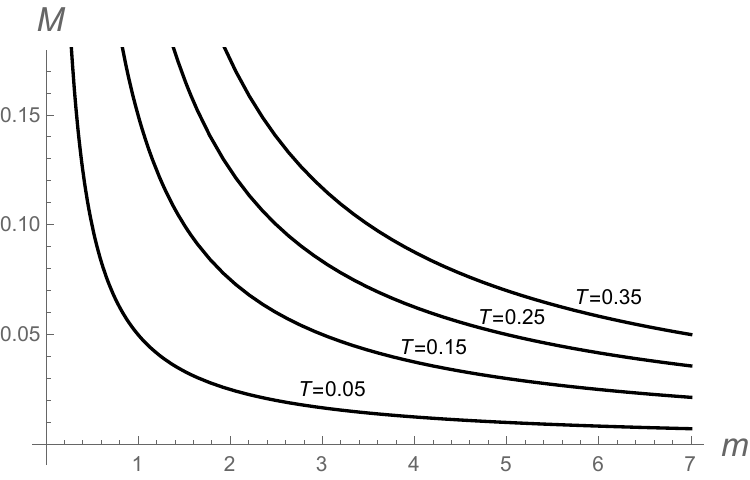}
	\end{center}
	\caption{\small The isothermal $M$-$m$ curves by (\ref{10}). }
\end{figure}

Now let us consider the behavior of the same quantity by the compact variable model 
(\ref{03}). The spectrum and the statistical mechanics of the model by (\ref{03}) are 
studied in \cite{spchfath,pvfath}. In particular, it is discussed in detail how the particle 
dynamics interpretation of the spin chain model as (\ref{03}) leads to the 
first-order phase transition. The one-particle partition function by the action (3),
using $L=2\pi R$, then is read
\begin{align}\label{13}
Z_1(\beta,\kappa  )=
\int_{-\pi/R}^{\pi/R} \prod_{n=0}^{N -1}\frac{dx_n}{\sqrt{2\pi}\, gR} 
\,\exp\left[\frac{1}{g^2}  \sum_{n=0}^{N-1} \left(\cos\frac{x_{n+1}- x_n}{R}-1\right)\right]
\end{align}
supplemented by the periodic condition $x_0=x_\beta$.
Thanks to the identity by the modified Bessel function of the first kind \cite{mattis}:
\begin{align}\label{14}
\exp[\kappa  \cos(y'- y)]=\sum_{s=-\infty}^\infty I_s(\kappa  ) \, e^{\mathrm{i}\,
s\,( y'- y)}
\end{align}
the above partition function can be converted to the following summation \cite{mattis,spchfath,pvfath}
\begin{align}\label{15}
Z_1(\beta,\kappa  )=\sum_{s=-\infty}^\infty 
\left( \frac{\sqrt{2\pi}}{g} \, e^{-1/g^2} I_s(1/g^2) \right)^N
\end{align}
by which, using $a N=\beta$ and the definition 
$Z=\sum_s e^{-\beta E_s}$, one reads the energy spectrum of the model \cite{mattis,spchfath,pvfath}
\begin{align}\label{16}
E_s(g)=-\frac{1}{a} \ln \left[\frac{\sqrt{2\pi}}{g} \, e^{-1/g^2} I_s(1/g^2)\right]
\end{align}
Using $I_0\geq I_s$, we see that the ground-state energy is by $s=0$. 
The property $I_s=I_{-s}$ also makes the spectrum doubly
degenerate for $s\neq 0$. The small coupling limit $g\ll 1$ is given by the asymptotic behavior \cite{spchfath,pvfath}
\begin{align}\label{17}
I_s(\kappa)\simeq \lim_{\kappa  \to\infty}
\frac{e^{\kappa }}{\sqrt{2\pi \kappa  }} \exp\left(-\frac{s^2}{2\kappa  }\right)
\end{align}
by which $~E_s\simeq  g^2 s^2/(2a)$,
matching the energy $E=p^2/(2m)$ of a free particle with 
momentum $p=s/R$ along the compact direction, and mass $m=a/(g^2R^2)$ by (\ref{07}). 
So in the limit $g^{-2} \to\infty$ the spectrum approaches to that of an ordinary particle as 
expected. In the strong coupling limit $g\to \infty$, using
$I_s(\kappa)\simeq ({\kappa}/ {2})^s/s!$ for $\kappa\ll 1$,
we have
\begin{align}\label{18}
E_s=(s+\frac{1}{2})\,\frac{\ln g^2}{a}+O(s\ln s) +O(g^{-2})
\end{align}
\begin{figure}[!ht]
	\begin{center}
		\includegraphics[scale=0.7]{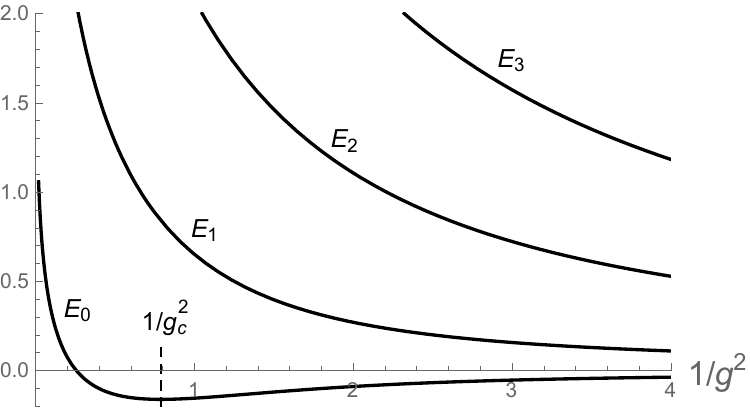}
	\end{center}
	\caption{\small The few lowest energies by (\ref{14}) versus $\kappa  $
		($E$ unit: $a^{-1}$) \cite{spchfath,pvfath}.}
\end{figure}

The remarkable fact about the spectrum of the model is the existence of a minimum in 
the ground state at $1/g_c^2=0.7902$ ($g_\mathrm{c}=1.125$) \cite{spchfath,pvfath}; see Fig.~2. 
The minimum in the spectra is specially important in connection with the phase transition. 
It is well known that the 1D spin systems with short range interactions do not exhibit phase transition. 
However, as the consequence of the particle dynamics interpretation, the present model 
exhibits a first-order phase transition \cite{spchfath,pvfath}. The nature of phase transition 
by the model can be studied based on the behavior of the Gibbs free energy \cite{spchfath,pvfath}. 
As the spectrum depends on $g$, by $m\propto 1/g^2$ and in analogy with (\ref{11}),
we define the thermodynamical conjugate variable of $1/g^2$ ($A_1=-T\ln Z_1$)
\begin{align}\label{19}
{M}(\beta,g^{-2}):= - \frac{\partial A_1}{\partial g^{-2}} 
\end{align}
which is also interpreted as the equation-of-state of the system.
The Gibbs free energy is then defined by 
\begin{align}\label{20}
G_1=A_1+g^{-2}  \,M
\end{align}
\begin{figure}[!ht]
	\begin{center}
		\includegraphics[scale=0.75]{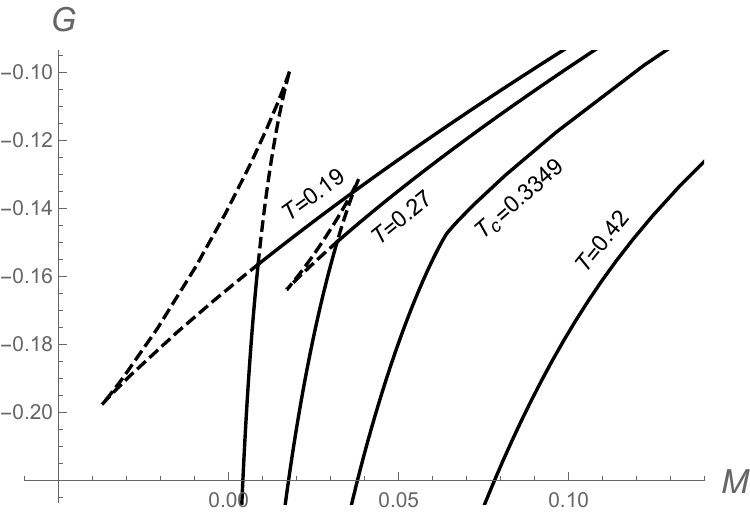}
	\end{center}
	\caption{\small The $G$-$M$ plots at four temperatures. The dashed pieces are
		not followed by the system due to the minimization of $G$ \cite{spchfath,pvfath}.}
\end{figure}

The isothermal $G$-$M$ plots are presented in Fig.~3. 
As seen, below the critical temperature $T_c=0.3349$ the plots develop cusps, 
at which the system follows the path with lower $G$ (solid-lines in Fig.~3), 
by the minimization of $G$ at equilibrium \cite{huang,stanley}. 
As the consequence, for $T<T_c$ there is a jump in first derivative 
of $\partial G/\partial M$, indicating that the phase transition is 
a first order one. It is apparent by now that the above phase structure 
is quite similar to the gas/liquid transition, for which $G$-$P$ plots 
show exactly the same behavior. In the similar way the equation-of-state 
(\ref{17}) should be modified by the so-called Maxwell construction for 
$P$-$V$ diagram, by which during isothermal condensation the pressure 
(here $M$) is fixed \cite{huang,stanley}. The results of the Maxwell 
construction for the present model are plotted as isothermal $M$-$1/g^2$ 
curves in Fig.~4. The flat part at $T_c$-curve corresponds to the critical values:
\begin{align}\label{21}
T_c=0.3349:~~1/g_*^2=1.4030,~~~ M_*=0.06419
\end{align}
corresponding to the coupling $g_*=0.8443$. 
\begin{figure}[!ht]
	\begin{center}
		\includegraphics[scale=.9]{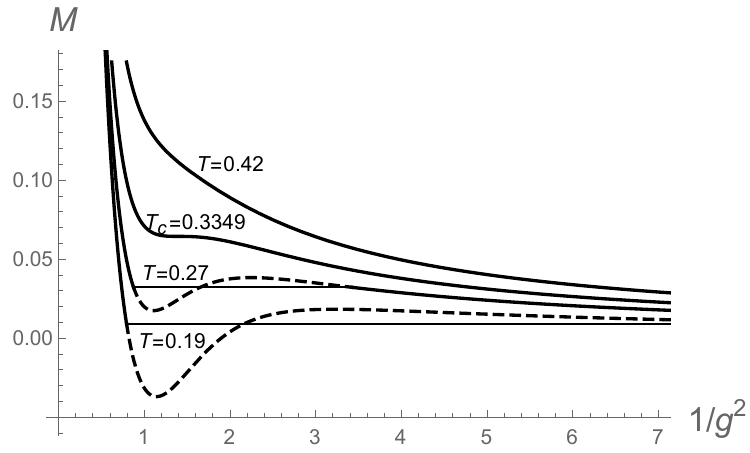}
	\end{center}
	\caption{\small The isothermal curves due to the Maxwell construction \cite{spchfath,pvfath}. The dashed parts in the upper plot represent the curve without the modification. In the lower plot the bold-line curve corresponds to the temperature $T_c=0.3349$.}
\end{figure}

The interesting fact about the equation-of-states modified by Maxwell construction is that $M$ always remains non-negative, that is $M\geq 0$.  This is specially important 
considering the expectation from (\ref{12}) as $M\propto \langle \dot{x}^2\rangle$.
The difference between the ordinary case and the present model 
is best understood comparing the isothermal 
curves by Figs.~1 \& 4. 
In both figures low and large masses are governed by vertical and horizontal 
asymptotes, respectively, although with different slopes. 
The main difference between the ordinary case and the model is due to the exhibited 
phase transition. In particular, by the present model and below the critical 
temperature $T_c$, the two asymptotes by large and small masses (large and 
small $g^{-2}$'s) are connected with a first order phase transition. 

The phase transition by the present model for particles with 
effective mass $m\propto 1/g^2$ can be matched to  
the proposed role of magnetic monopoles in connection with 
the two phases of the U(1) Abelian gauge theory. 
As mentioned earlier, according to the scenario based on the dual Meissner effect, it is the motion of monopoles in the presence of source electric charges that determines the electric fluxes' shape. By the present model 
below $T_c$ the behavior of $M\propto \langle v^2\rangle$ at weak and strong coupling limits are connected by a first order phase transition. 
\begin{figure}[!ht]
	\begin{center}
		\includegraphics[scale=.9]{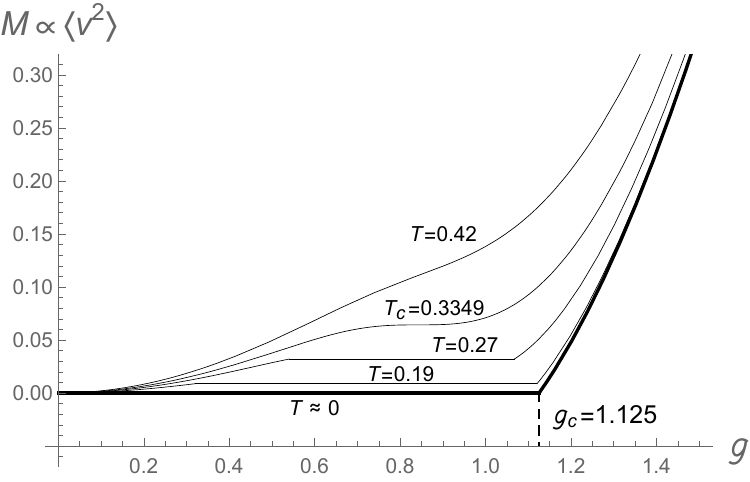}
	\end{center}
	\caption{The isothermal $M$-$g$ plots \cite{spchfath}.}
\end{figure}

The behavior of system at low temperatures
is of particular interest. In the limit $T\to 0$, due to the Maxwell construction, we have $M=0$ for $g<g_{c}=1.125$; Fig.~5. 
This behavior due to the discontinuous nature of first order phase transition is 
to be compared with (\ref{12}), by which $M$ increases gradually by lowering the 
mass at constant $T$.
Hence, by the role proposed for the motion of monopoles, 
at very low temperatures and below $g_{c}$ 
the Coulomb phase stays unrivaled with $\langle v^2 \rangle \approx 0$.
On the other hand, exhibiting a high-slope increase of $\langle v^2 \rangle$ at $g_{c}$, the confined phase at low temperatures should correspond to 
$g > g_{c}$. This picture and specially the value of critical coupling constant suggested by lattice
simulations $g_{c}\approx 1$ are in agreement with theoretical and numerical studies mentioned earlier. 

\vspace{2mm}
\textbf{Acknowledgement}: 
This work is supported by the Research Council of Alzahra University.



\end{document}